\documentclass[
superscriptaddress,
reprint,
 amsmath,amssymb,
 aps,
prl,
]{revtex4-1}

\usepackage{graphicx}% Include figure files
\usepackage{dcolumn}% Align table columns on decimal point
\usepackage{bm}% bold math
\usepackage{amsmath}
\usepackage{amsfonts}
\usepackage{amssymb}
\usepackage{dsfont}
\usepackage{color}
\usepackage{xcolor}
\usepackage{siunitx}
\usepackage[colorlinks=true,
    linkcolor=blue,
    filecolor=blue,      
    urlcolor=blue,
    citecolor = blue]{hyperref}
\usepackage{braket}
\usepackage{mathtools}
\usepackage[mathlines]{lineno}% Enable numbering of text and display math
\usepackage[normalem]{ulem}
\usepackage{scalerel}
\usepackage{ bbold }
\usepackage{inputenc}
\usepackage[T1]{fontenc}
\usepackage{graphicx}
\usepackage{siunitx}
\usepackage[caption=false]{subfig}

\graphicspath{{./figures/}}

\begin{document}

\title{Chirped pulse control over the melting of superconductors} 

\date{\today}
\author{Maria Recasens}
\affiliation{ICFO - Institut de Ciencies Fotoniques, The Barcelona Institute of Science and Technology, Av. Carl Friedrich Gauss 3, 08860 Castelldefels (Barcelona), Spain}
\email{maria.recasens@icfo.eu}

\author{Valentin Kasper}
\affiliation{ICFO - Institut de Ciencies Fotoniques, The Barcelona Institute of Science and Technology, Av. Carl Friedrich Gauss 3, 08860 Castelldefels (Barcelona), Spain}
%\affiliation{Department of Physics, Harvard University, Cambridge, MA, 02138, USA}
\email{valentin.kasper@icfo.eu}

\author{Maciej Lewenstein}
\affiliation{ICFO - Institut de Ciencies Fotoniques, The Barcelona Institute of Science and Technology, Av. Carl Friedrich Gauss 3, 08860 Castelldefels (Barcelona), Spain}
\affiliation{ICREA, Pg. Lluis Companys 23, 08010 Barcelona, Spain}
\author{Allan S. Johnson}
\affiliation{IMDEA Nanoscience, Calle Faraday 9, 28023 Madrid, Spain}
\email{allan.johnson@imdea.org}

\begin{abstract}
Strong field terahertz pulses are increasingly used to excite and control quantum materials at the ultrafast timescale. They have found widespread application by enabling direct addressing of the superconducting gap or Josephson resonances and are essential in Higgs spectroscopy. Large non-linear optical signals can be induced by the strong coupling of the THz and superconducting degrees of freedom. However, far less attention has been paid to the strong bi-directional coupling between field and material this implies. Here, we use the framework of the time-dependent Ginzburg-Landau equations to study the full field and material evolution of a superconductor driven by strong field terahertz pulses. We find that at high field strengths, the backreaction of the superconductor induces large changes to the driving pulse, which in turn leads to a runaway melting of the superconducting condensate. This results in a surprisingly large sensitivity to the initial driving pulse chirp, enabling these purely dynamical changes to result in order of magnitude different levels of melting. We also find large-scale spectral shifting of the driving pulse to occur in just a few hundred nanometers of propagation through a superconductor. We attribute these effects to an inverse plasma redshift, in which the driving field breaks Cooper pairs and decreases the free-electron mobility, analogous to reducing the density of a plasma. 
\end{abstract}
\maketitle

\emph{Introduction.} Terahertz (THz) pulses have recently emerged as a powerful tool for exciting and controlling quantum materials. The low photon energies allow the THz radiation to couple the collective excitations efficiently in various systems. Furthermore, with the advent of high-field pulses with peak electric field strength of hundreds of kilovolts to megavolts per centimeter, it has become possible to drive quantum materials far out of equilibrium~\cite{yang2023terahertz}. THz driving has been particularly successfully applied to superconductors with various low-energy collective modes. Examples include the use of THz driving to observe hidden stripe phases, parametrically amplify superconducting fluctuations, and coherently drive Higgs modes, leading to new forms of nonlinear spectroscopy, emergent phases, and control of superconductivity~\cite{cavalleri_photo-induced_2018, budden_evidence_2021, Shimano2020}.

In many of these cases, it has been noted that the nonlinear response of the superconductor to the THz radiation is exceptionally large, with cross sections far greater than can be found in conventional materials~\cite{matsunaga_light-induced_2014, chu_phase-resolved_2020}. This is true both when the THz is resonant to particular collective modes, but also works off-resonance when the primary coupling is directly to the superconducting condensate~\cite{cea_nonlinear_2016,robson_giant_2017}. Especially, in the latter case, the light field can drive large time-dependent supercurrents. Numerous studies and models have been made to explain these large non-linear effects with focus upon the accurate treatment of the superconducting response~\cite{cea_nonlinear_2016, Shimano2020, demsar_non-equilibrium_2020}. However, at high-fields it can be important to model the entire system of light and superconductor~\cite{robson2017giant,robson_giant_2018}, i.e., the back-reaction of the superconductor on the THz field. This back-reaction can drastically modify the response of the system, and the ability to drive and control superconductors with light selectively~\cite{cea_nonlinear_2016}. As larger field strength THz pulses become available~\cite{doi:10.1021/acsphotonics.2c01336}, this backreaction can be expected to be even more significant.

Here we present work modeling the impact of the superconducting backreaction on an intense THz driving field and how this strongly modifies the overall system dynamics. We directly model and solve a one-dimensional system to study the response of a superconducting condensate and the propagation of the THz driving laser field. Particularly, we treat the superconducting response via the phenomenological time-dependent Ginzburg Landau (TDGL) equations~\cite{robson_giant_2017} and the field propagation by the full Maxwell equations. We focus on the effect of the initial chirp of the pulse in the interaction~\cite{warren_coherent_1993}. Our results demonstrate a large dependence of the system on the initial pulse chirp at high THz field strengths, with realistic chirp variations leading to differences in the superconducting melting up to one order of magnitude. This dependence is attributed to a spectral shift in the driving pulse occurring within a range of a few hundred nanometers. We associate this spectral shift with the breaking of Cooper pairs caused by the pulse, giving rise to an inverse plasma redshift effect~\cite{BLOEMBERGEN1973285}. The substantial nonlinear response observed in the superconductor has the potential to improve control and diagnostics of superconductivity.

%-------------------------------------------------------------------------------------------
\section{Model}
%\subsection{Ginzburg-Landau approach}
%What it is + Motivation TDGL equation

A wide variety of models have been used previously to describe the dynamics of superconductors out of equilibrium. These include microscopic models like the time-dependent Gutzwiller variational approach or time-dependent Dynamical Mean Field Theory, see~\cite{giannetti2016ultrafast} for a review. On the other hand phenomenological models, such as the two-temperature model or the Rothwarf-Taylor model~\cite{coslovich2011evidence}, can return qualitative understanding about the melting process but generally do not directly treat the coupling between laser pulses and superconductors. 

For our simulations, we use the time-dependent Ginzburg Landau theory coupled to the Maxwell equations. The TDGL equation describes the dynamics of the superconductor order parameter $\psi$ and the Maxwell equations the propagation of the EM field. The TDGL equation was derived by Schmidt~\cite{schmid_time_1966}, and later Gor'kov and Eliashberg showed that it could also be derived from the BCS theory in the case of gapless superconductors ~\cite{Gorkov_microscopic}. The order parameter $\psi$ can be directly related to the Cooper pair density $n_c$ via $n_c = |\psi|^2 $ through BSC theory. Critically, the TDGL makes concrete and quantitative predictions for a wide variety of dynamical behaviors, such as the dynamics of vortices in type II superconductors~\cite{dorsey_vortex_1992,machida_direct_1993} or of fluctuations in the conductivity above $T_{c}$~\cite{schmidt_onset_1968}. TDGL has previously been applied to studying THz driving of superconductors, where it was found that the TDGL could semi-quantitatively describe the giant Kerr response in superconductors~\cite{robson_giant_2017,robson_giant_2018}. We briefly note that an alternative approach to treat the complete field+superconductor system would be models based on the Boguliobov-de Gennes Hamiltonian ~\cite{vaswani_light_2021, luo2023quantum, PhysRevB.102.054517}.

The TDGL equation~\cite{gropp_numerical_1996} is 
\begin{align}
\frac{\hbar^2}{2m^{*}D}\frac{\partial \psi}{\partial t} = \frac{\partial f}{\partial \psi^*}\, ,
\label{eq_relaxation}
\end{align}
where $D$ is the diffusion coefficient, $m^{*}$ is the effective mass of the Cooper pair and $f$ is the free energy, where we already 
implicitly assume the temporal gauge $\phi =0$. 
In the Ginzburg-Landau equation for superconductivity, the free energy is given by
\begin{align}
f = f_{n} + a |\psi|^2 + \frac{b}{2} |\psi|^4 + \frac{1}{2m^{*}} |(-i\hbar \nabla -e^{*}\textbf{A})\psi|^2 \, ,
\label{eq_freeGL}
\end{align}
where $a=a_{0}\left(T-T_{c}\right), a_{0}$ and $b$ are constants of the material, $e^{*}$ and $m^{*}$ are the effective charge and mass of the Cooper pairs, and $\boldsymbol{A}$ is the vector potential that represents the external light field. 
The parameters $a_{0}$ and $b$ can be determined through their relationships to the experimentally measurable London penetration depth $\lambda$ and coherence length $\xi$, which are defined as;
\begin{subequations}
\begin{align}
\xi (T) &= \sqrt{\hbar^2/(2m^{*}|a(T)|)}, \\
\label{eq_coherencelengh}
 \lambda(T) &= \sqrt{m^{*}/(4 \mu_0 e^2 |\psi_0|^2)} \, , \\
|\psi_{0}|^2 &= -a_{0}/b,
\label{eq_psi0} \\
\tau &= \xi^2/D \label{eq:relaxation_time}
\end{align}    \label{eq:abbreviations}
\end{subequations}
\noindent where $|\psi_{0}|^2$ is the solution of the TDGL equation in the case of no spatial dependence and no electromagnetic field applied. The Ginzburg–Landau parameter, $\kappa=\lambda/\xi$, can also be obtained through these quantities.

By inserting the free energy of Eq.~\eqref{eq_freeGL} in Eq.~\eqref{eq_relaxation}, we obtain
\begin{align}
\frac{\hbar^2}{2m^{*}D}\frac{\partial \psi}{\partial t} = \frac{1}{2m^*}(-i\hbar \partial_{j} -2eA_{j})^2\psi-(a + b|\psi|^{2})\psi.
\label{eq_TDGLcomplet}
\end{align}
Further, we rewrite the TDGL using the abbreviation introduced in Eq.~\eqref{eq:abbreviations} leading to 
\begin{align}
\tau\partial_t\psi = \xi^{2}\left(-i \partial_{j} -\frac{2e}{\hbar} A_{j}\right)^2\psi+\left(1-\frac{|\psi|^{2}}{|\psi_{0}|^2}\right)\psi.
\label{eq_TDGL}
\end{align}

%remarks about TDGL equations form of the TDGL equation can be derived directly from the BCS theory applied to gapless superconductors since Eq.~\eqref{eq_relaxation} is phenomenological and valid for cases where $\psi$ decays back to its equilibrium value, the validity of Eq.~\eqref{eq_TDGLcomplet} is not restrained these cases. For example, in the case of finite-gap superconductors, a microscopic derivation of the theory that obtains a similar version of the TDGL equations has also been proposed~\cite{guljan_shortcut_2020}. \par

%Where do we talk about the dimensions leading to another TDGL expression? 

%Supercurrent 
To describe the superconductor interaction with the EM pulse, it is also necessary to include in the model the supercurrent equation, i.e., the dissipation-less current,
\begin{align}
    J_s = \frac{e^*}{m^{*}} \text{Re}[\psi^*(-i\hbar \nabla - e^* \textbf{A})\psi]\, ,
    \label{eq_supercurrent}
\end{align}
where $\psi^*$ is the complex conjugate of $\psi$ and $\text{Re}$ denotes the real part~\cite{tinkham_introduction_2015}. \par

Furthermore, to fully describe the coupled system, we also need to include the response of normal-state materials. Here, we use the Drude model to consider the remaining free electrons of the superconductor not bound into Cooper pairs. The evolution of the electric field is, as a result, described by
\begin{subequations}
\begin{align}
 \partial_t E_x(y,t) &= \epsilon^{-1}_0\partial_y H_z(y,t) + \epsilon^{-1}_0 [J_d (t) + J_s(y,t)],
 \label{eq_dEdt} \\
\partial_t H_z(y,t) &= \mu^{-1}_{0}\partial_y E_x(y,t) \, ,
\label{eq_dHdt} 
\end{align} \label{eq:Maxwell}
\end{subequations}
where $\epsilon_0$ is the permittivity of free space and $J_d$ is the Drude current, whose dynamics are described by 
\begin{align}
    \partial_t J_{d}(t) &= - \gamma J_d(t) + \omega_p^2\epsilon_0E_x(y,t) \, ,
\label{eq_drudecurrent}  
\end{align}
where $\gamma$ is a damping factor and $\omega_p$ the plasma frequency. Thus, the entire system of equations we solve to describe the material response and field evolution is given by Eq.~\eqref{eq_TDGL}~-~\eqref{eq_drudecurrent}. The TDGL theory is valid within certain limits, specifically when the frequency of the time variation of $\psi$ is greater than the gap frequency as discussed in~\cite{abrahams1966time, tinkham_introduction_2015}. 

%-------------------------------------------------------------------------------------------
\subsection{Geometry and setup}
\begin{figure} [h]
    \includegraphics[width=0.5\textwidth]{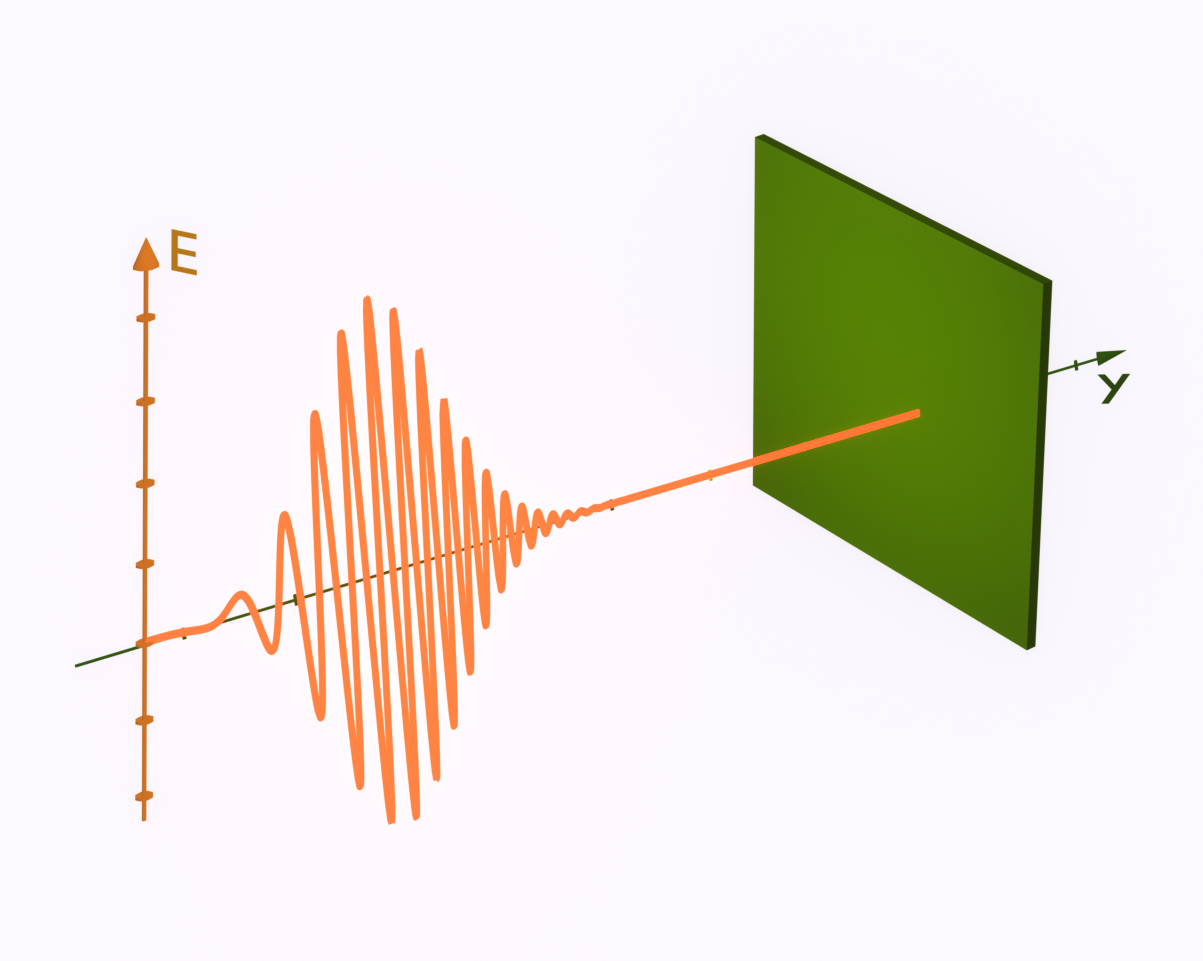}
    \caption{\textbf{Simulation setup}. We consider a thin, infinite two-dimensional superconducting plate (green) with a chirped probing pulse (orange). }
  \label{Fig_sketchgeometry}
\end{figure}

To investigate the back reaction of a superconductor on a THz field, we simulate the case of a high-field THz pulse incident on bulk material. Specifically, we focus on the effect of the initial chirp of the pulse, as introducing a frequency chirp modifies the pulse without affecting the total energy of the pulse. In particular, we model the scenario of a THz pulse hitting a superconducting thin film as illustrated in Fig.~\ref{Fig_sketchgeometry}, allowing us to describe the system in one spatial dimension.  For the explained geometry of the simulation, the order parameter $\psi$ can hence be treated as only dependent on $y$. 

\subsection{Pulse and Material}
The dynamics of a superconductor driven by a THz pulse will depend on the parameters describing the geometry, the incident pulse, and the material properties. Concerning the incoming pulse, we mainly focus on the pulse chirp. Tuning the chirp does not change the total input energy, which gives a unique look into the dynamic response of the system. 

Regarding the driving THz field, we start from a simple Gaussian pulse with an electric field given by 
\begin{align}
    E(t) = E_0e^{-(t-t_0)^2/t_g^2}\cos{[\omega_0(t-t_0)]} \, , 
\label{eq_Gaussian pulse}
\end{align}
where $E_0$ is the electric field strength, $t_g$ determines the width of the envelope in the time domain, $\omega_0$ is the angular photon frequency, and $t_0$ is the center of the pulse in the time domain. \par
We include the chirp by multiplying the Fourier transform of the pulse given in Eq.~\eqref{eq_Gaussian pulse} with 
\begin{align}
  e^{-i[a(2\pi f-\omega_0) + b(2\pi f-\omega_0)^2 
                                  + d(2\pi f-\omega_0)^3 + ...]}
\label{eq_chirp}
\end{align}
and then Fourier transforming back to the time domain, where $a$ quantifies the scalar phase shift and $b$ the linear chirp. This pure phase addition stretches the pulse in the time domain but does not change the total energy or the spectral content. Instead, it modifies the arrival time of the different spectral components. To obtain a chirp of order $n - 1$, we add a term proportional to $(2\pi f-\omega_0)^n$ in the exponent of Eq.~\eqref{eq_chirp}. 

Concerning the initial pulse, we explore a range of field strengths spanning one order of magnitude, from $\SI{0.9}{\mega\volt\per\centi\meter}$ to $\SI{9}{\mega\volt\per\centi\meter}$ for the unchirped pulses. These fields are high but technically feasible~\cite{doi:10.1021/acsphotonics.2c01336}. Further, we note using thinner samples allows us to imitate lower field strengths in certain circumstances. We consider photon frequencies ranging from the THz to the mid-infrared domain with $t_g=\SI{132}{\femto\second}$ and $\omega_0=\SI{4.8}{\tera\hertz}$. As for the frequency chirp, we introduce a linear chirp, keeping the parameters $a=c = 0$.

Motivated by experiment, we consider a $\SI{310}{\nano\meter}$ thick film of lead, with a critical temperature of $T_c$ of $\SI{7.2}{K}$, see~\cite{Wesche2017}. The material choice determines the TDGL parameters. For a BCS superconductors superconductor, $m^*$ and $e^*$ take straightforward values related to the Cooper-pair structure, $m^*=2m_e$ and $e^*=2e$~\cite{Cabrera1989, Kov1959}. For lead, the remaining parameters take the following values $\xi=\SI{82}{\nano\meter}$, $\kappa=0.48$~\cite{Wesche2017} and $\rho=\SI{3.84e-10}{\ohm\meter}$. The simulations are initialized with the system in equilibrium at $T = \SI{0}{K}$.

\section{Results}
We calculate the propagation of the pulse and the dynamics of the material by numerically integrating the system of equations Eq.~\eqref{eq_TDGL} to Eq.~\eqref{eq_drudecurrent}. At the boundaries of the material, we enforce $\psi = 0$ and generate the propagating pulse using a source far away from the material. To avoid reflections of the EM pulse at the boundaries of the simulation, we added perfectly matched layers far away from the source and the material. Finally, we discretize space using finite differences and solve the time evolution with a 4th-order Runge-Kutta solver. In the following subsections, we analyze our simulation results, focusing on the influence of the chirp. We analyze the effect of the sign of the chirp, the response of the superconducting order parameter concerning the chirp, and the response of the electromagnetic field.

\begin{figure*} 

\subfloat[\label{subfig:a}]{%
  \includegraphics[width=1.0\textwidth]{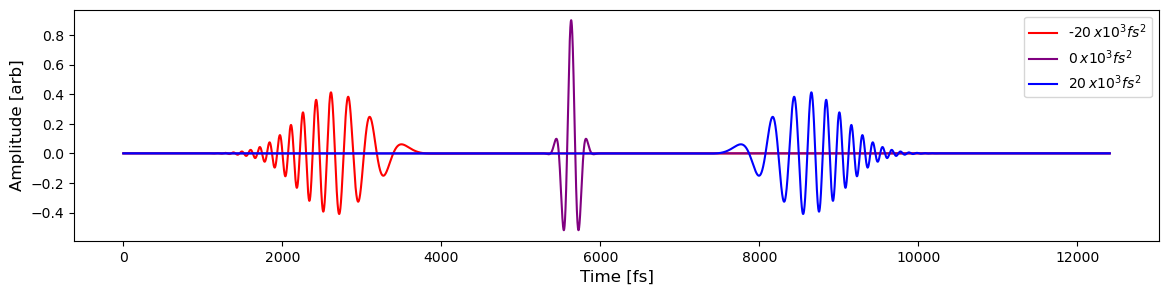}
  \label{fig:a} 
}\hfill

\subfloat[\label{subfig:b}]{%
  \includegraphics[width=\columnwidth]{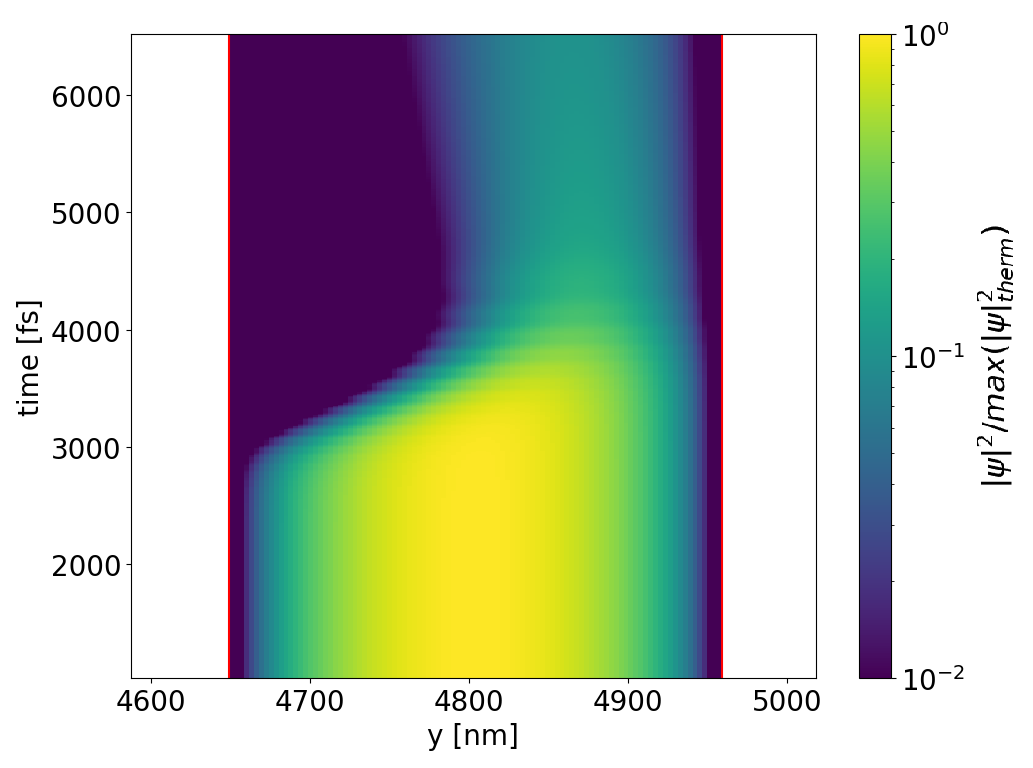}
  \label{fig:b} 
}
\subfloat[\label{subfig:c}]{%
  \includegraphics[width=\columnwidth]{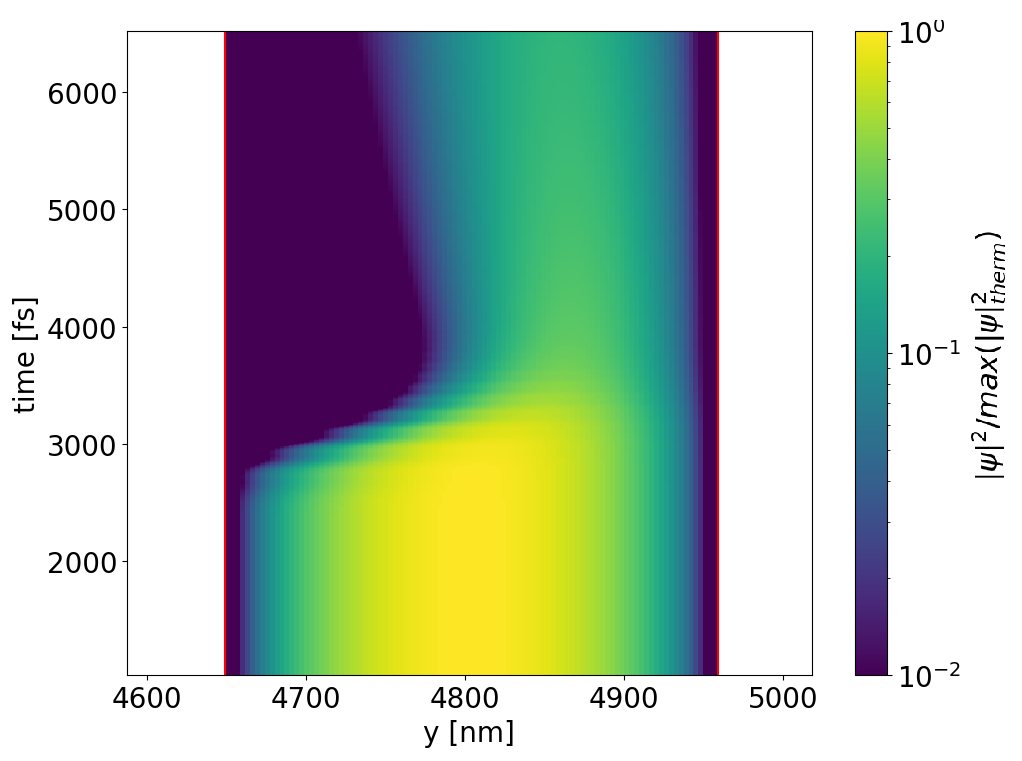}
  \label{fig:c} 
}\hfill

\subfloat[\label{subfig:d}]{%
  \includegraphics[width=\columnwidth]{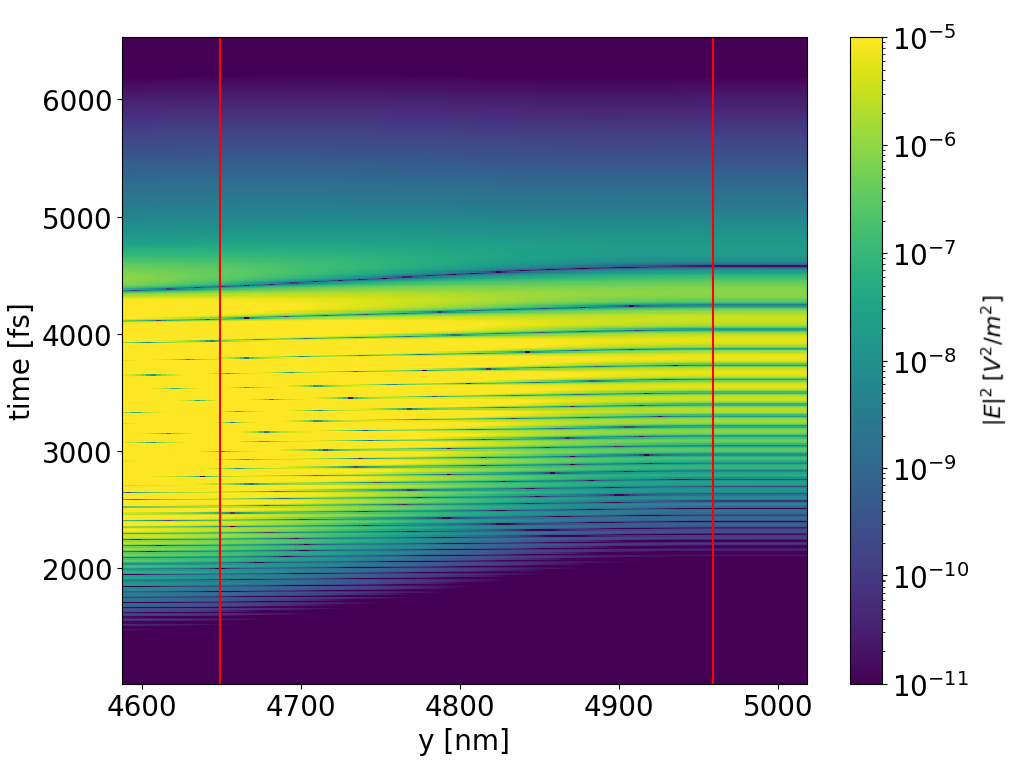}
  \label{fig:d} 
}
\subfloat[\label{subfig:e}]{%
  \includegraphics[width=\columnwidth]{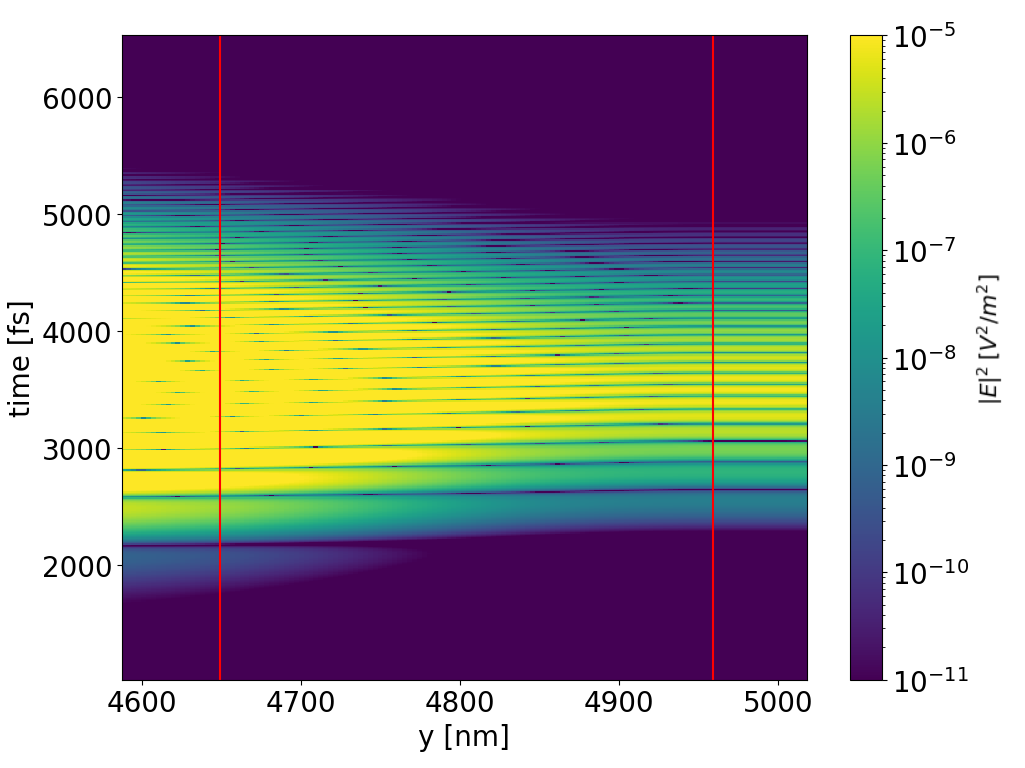}
  \label{fig:e} 
}\hfill
\caption{\textbf{Pulse sequence and spatio-temporal distribution of order parameter} Interaction between the superconductor and the electromagnetic pulse for a peak field strength of $\SI{8}{\mega\volt\per\centi\meter}$ for the unchirped pulse. The red lines represent the boundaries of the superconductor. In subfigure (a) we show the electric field of EM pulses with different linear frequency chirps.  Further, in subfigure (b) and (c) the order parameter as a function of time and space normalized by the maximum value of the initial thermal state, $|\psi|^2/|\psi_{\text{max}}|^2$, where $|\psi_{\text{max}}|^2 = \SI{3.8517}{\per\nano\meter\cubed}$. In subfigures (d) and (e) we show the time and space evolution of the electric field amplitude squared, $E^2$ in $\SI{}{\volt\squared\per\meter\squared}$. The explicit value of the chirp is $b = - \SI{20e3}{\femto\second\squared}$ in (b) and (d), and $b = \SI{20e3}{\femto\second\squared}$ in subfigure (c) and (e).}
\label{Fig_time_space}
\end{figure*}

\begin{figure*}[t!]
\subfloat[  \label{2fig:a} ]{%
  \includegraphics[width=0.97\columnwidth]{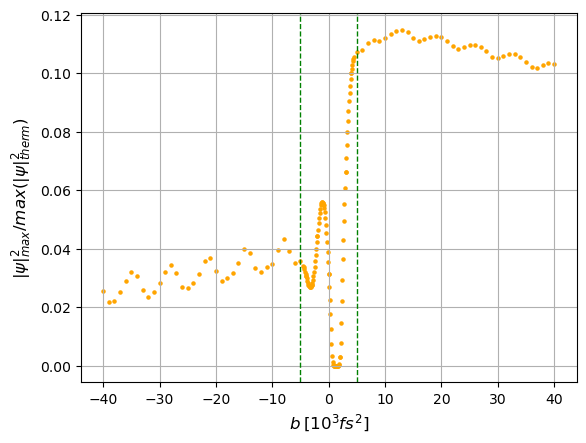}
}
\subfloat[\label{2fig:b} ]{%
  \includegraphics[width=\columnwidth]{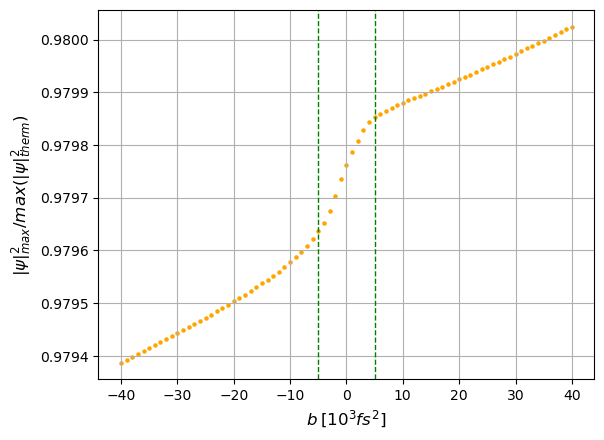}
}\hfill
\caption{\textbf{Tuning order parameter with linear chirp.} The maximum value of the order parameter after the pulse has passed the superconductor, at $t = \SI{12}{\pico\second}$ for two different initial field strengths. The value of the order parameter is normalized by the maximum value of the initial state, $|\psi_{\text{max}}|^2/\text{max}(|\psi_{\text{therm}}|^2)$ with $\text{max}(|\psi_{\text{therm}}|^2) = \SI{3.8517 }{\per\nano\meter\cubed}$. Green vertical lines denote where the chirp is of magnitude $\SI{5e3}{\femto\second\squared}$. In subfigures (a) and (b), we use a peak field strength of $\SI{9}{\mega\volt\per\centi\meter}$ and $\SI{0.9}{\mega\volt\per\centi\meter}$ respectively for $b=0$.}
  \label{Fig_maxphi_chirp}
\end{figure*}

\subsection{Positive chirp - negative chirp}
We simulate the interaction between the superconductor and the pulse for two different pulses, one with positive and one with negative chirp. Fig.~\ref{fig:a} illustrates the electric field of pulses with chirps of $\SI{20e3}{\femto\second\squared}$, $\SI{0}{\femto\second\squared}$ and $\SI{-20e3}{\femto\second\squared}$. The high-frequency components arrive first for positive chirps, while for negative chirps we observe the opposite.

In Fig.~\ref{Fig_time_space}, we study the spatial and temporal results of the described simulations. Specifically, Fig.~\ref{fig:b} and Fig.~\ref{fig:d} show the results for the negative linear chirp, whereas Fig.~\ref{fig:c} and Fig.~\ref{fig:e} depict the results for the positive linear chirp. By comparing the left and right columns, we note that the two scenarios show significant qualitative differences in the evolution of the superconductor. Specifically, we observe a more significant suppression of the order parameter for negative chirps, and the suppression occurs at earlier times for the positive chirp case. In addition, the figures Fig.~\ref{fig:d} and Fig.~\ref{fig:e} show the evolution of the corresponding EM pulse inside the superconductor. For the pulse with positive chirp, see Fig.~\ref{fig:e}, we observe that low and high frequencies are similarly suppressed inside the superconductor, whereas for the negative chirp pulse, see Fig.~\ref{fig:d}, high frequencies are more suppressed than lower frequencies. These differences indicate tuning the response of the material by the sign of the chirp. 

\subsection{Suppression of superconductivity using the chirp}
To gain more insight into the chirp dependence of the dynamics, we systematically vary the chirp and quantify the extinction of the order parameter by studying the maximum value of $\psi_{\text{max}}$ inside the superconductor at the end of the simulation, specifically 
$\psi_{\text{max}}$ at $\SI{12}{\pico\second}$. The results of this study are given in Fig.~\ref{Fig_maxphi_chirp} where we show $\psi_{\text{max}}$ as a function of the linear chirp of the source pulse for two different field strengths one order of magnitude apart.

% \begin{figure*} [t!]

%  \begin{subfigure}{0.49\textwidth}
%  \subcaption{   \hfill{} }
%     \label{2fig:a}
%      \includegraphics[width=0.97\textwidth]{phimax_tg10_0.9.png}

%  \end{subfigure}
%  \hfill
%  \begin{subfigure}{0.49\textwidth}
% \subcaption{   \hfill{} }
%      \includegraphics[width=\textwidth]{0.09_phimax_tg10.png}

%      \label{2fig:b}
%  \end{subfigure}

% \caption{\textbf{Maximum value of the order parameter of the superconductor as a function of linear chirp.}  \textit{(a),(b):} Maximum value of the order parameter after the pulse has gone through the superconductor, at $t = 12 ps$. The value is normalized by the maximum value of the initial state, $\frac{|\psi|^2_{\text{max}}}{max(|\psi|^2_{\text{therm}})}$, where $max(|\psi|^2_{\text{therm}}) = 3.8517 \frac{1}{nm^3}$. Green vertical lines denote where the chirp is of magnitude $|$\num{5e3}$| \: fs^2$. \textit{(a):} For a peak field strength of 9 $\frac{MV}{cm}$ at b=0.  \textit{(b):} For a peak field strength of 0.9$\frac{MV}{cm}$ at b=0.}
%   \label{Fig_maxphi_chirp}

% \end{figure*}

In both plots, for chirps with an absolute value larger than $\SI{5e3}{\femto\second\squared}$, we can see that negatively chirped pulses suppress the order parameter more than positively chirped pulses. Thus, negative chirped pulses more effectively destroy superconductivity than positive chirped ones. As we discuss in the following section, this marked difference arises from the dependence of the backreaction of the superconductor on the sign of the chirp. This result holds for an extensive range of field strengths, although the magnitude of the effect changes significantly. In addition to this effect, there are clear oscillations in $\psi_{\text{max}}$ with chirp, along with exceptionally high suppression of the amplitude for chirps lower than $\SI{5e3}{\femto\second\squared}$ in the high field case.

\subsection{Optical back-reaction}

\begin{figure}
 \includegraphics[width=\columnwidth]{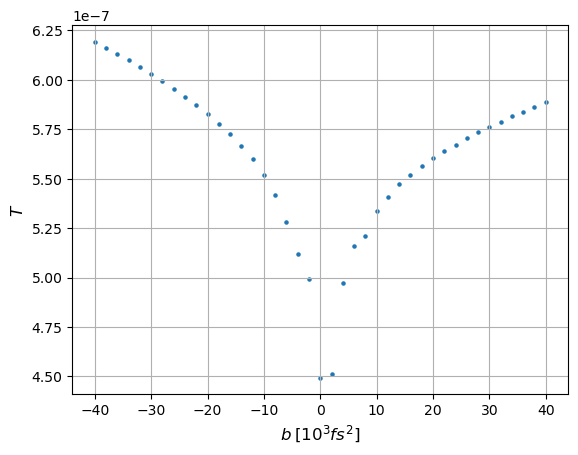}

    \caption{\textbf{Transmission coefficient:} For different chirps, we depict the total transmitted energy after the material normalized by the total energy of the source pulse for a peak field strength of $\SI{9}{\mega\volt\per\centi\meter}$ at $b=0$. We observe a clear minimal transmission coefficient close to zero chirp, and observe asymmetric tuning of of the transmission for positive and negative chirps.}
  \label{Fig_integral_intensity}
\end{figure}

To understand the origin of these changes in suppression, we first examine the transmission of the pulse as a function of chirp, as shown in Fig.~\ref{Fig_integral_intensity}, focusing on the high field case where the effects are more dramatic. We immediately note that the changes observed in Fig.~\ref{Fig_maxphi_chirp} do not map directly to changes in transmission; the asymmetry between positive and negative chirps is small, and there are no oscillations. However, the absorption increases dramatically for small chirp values at the highest pulse intensity. This behavior does not follow a simple power law of the intensity. It thus suggests that for low-chirp, high-intensity pulses, additional non-linear melting processes of the condensate become significant. We have verified this assertion by performing additional scans under zero chirp conditions in which we keep the total energy in the pulse constant but change $t_g$, leading to a decrease in peak field, which indeed confirms a nonlinear melting process occurs at high field strengths, see Supplementary Fig.~\ref{Fig_suppelmentary}.

\begin{figure*}
 \includegraphics[width=0.8\textwidth]{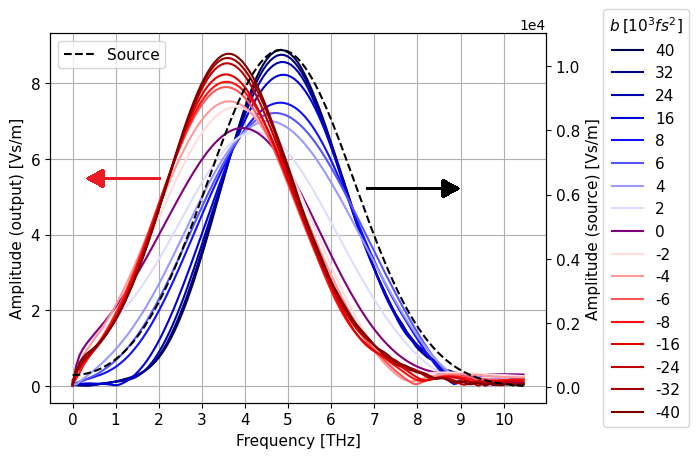}  
  
  \caption{\textbf{Spectrum of the laser pulse after interaction.} Spectra of the pulse after the material for a field strength of $\SI{9}{\mega\volt\per\centi\meter}$. Red solid lines correspond to the spectra of the pulses with negative chirps, purple solid lines to the spectra of the Gaussian pulse without chirp, and blue solid lines to the spectra of the positive chirped pulses. The dashed black line corresponds to the input spectra of the source, which is the same for all chirps.  }

  \label{Fig_spectras}
\end{figure*}

As changes in total absorption cannot explain the strongly asymmetric melting process nor the oscillations, we next examine the backreaction of the superconductor on the spectrum of the pulse. Fig.~\ref{Fig_spectras} displays the spectra of the source pulse and the output pulse spectrum for a range of chirps. Solid lines correspond to the spectra of pulses after going through the material, and the dashed black line is the spectra of the source pulses. 
We see a clear spectra shift towards lower frequencies for negative chirps lower than $\SI{-5e3}{\femto\second\squared}$. On the other hand, for positive chirps higher than $\SI{5e3}{\femto\second\squared}$, we observe no shift and an asymmetric narrowing of the spectra. This offers a partial explanation for the behavior of the superconductor; low-frequencies couple more strongly to the condensate~\cite{robson_giant_2017} and lead to a non-thermal suppression of the superconducting condensate, and hence this giant redshift enhances the non-thermal melting process. A redshift of this type has previously been predicted for THz-driven superconductors, but only under the zero-chirp condition where we see a significantly smaller redshift~\cite{robson_giant_2018}.

Finally, we discuss the origin of the oscillations in $\psi_{\text{max}}$ in Fig.~\ref{Fig_maxphi_chirp}. The magnitude of these oscillations increases with increasing $E_0$ magnitude. However, their period remains constant, and they do not appear in either the transmission or spectral shift. Changing the pulse central frequency $\omega_0$ does not affect the oscillation period either. Modifying the material thickness alters the period of the oscillations, suggesting that they can be attributed to the finite-size effects of the superconducting film. 

\subsection{Asymmetric spectral broadening and melting}
Previous work~\cite{robson2018giant} related the melting of superconductivity to a giant asymmetric third-order non-linear optical response.  Here, we offer an intuitive explanation of this effect and how this leads to strong chirp sensitivity of the melting process. A driven material with a time-varying refractive index modulates the phase of the driving field, leading to new frequencies, i.e., self-phase modulation ~\cite{Boyd2008}. Typically, the refractive index follows a simple rule $n(t)=n_0+3\chi^{(3)}I(t)/(4n_0^2\epsilon_0c)$, where $n_0$ is the static refractive index, $\chi^{(3)}$ is the third-order non-linear susceptibility and $I(t)$ is the intensity of the driving pulse. Because $n(t)$ then follows the temporal behavior of $I(t)$, there is a region where $n(t)$ increases, then decreases. For a positive $\chi^{(3)}$, which describes most materials, rising intensities lead to an increase in refractive index and a redshift, while falling intensities lead to a blueshift~\cite{Boyd2008}. 

In a superconductor, however, the leading contributions to $\chi^{(3)}$ come from the melting of the superconducting condensate~\cite{robson_giant_2017}. In this case, $n(t)$ does not follow $I(t)$ as the pulse can only melt the superconductor, and the evolution of $n(t)$ is monotonic. This is conceptually very close to the idea of the plasma-blue shift: in a gas ultrafast laser ionization generates free carriers, leading to the formation of a plasma and a blue shift in the driving laser frequency because of the associated time-dependent decrease in the refractive index~\cite{BLOEMBERGEN1973285,giulietti2015self}. However, in a superconductor, the inverse effect is at play; by breaking Cooper pairs (effectively melting a high-mobility plasma), we dynamically reduce the effective free-carrier response, leading to a redshift. Robson \emph{et al.}~\cite{robson2017giant}, derived an approximate expression for $\chi^{(3)}$ in superconductors
\begin{align}
\chi^{(3)} = - \frac{128k_b^2 (T-T_c)^2 e^4}{\epsilon_0 \beta \omega_0^6 \pi^2 \hbar^2 m_e^2},
\label{eq nonlinear electric susceptibility}
\end{align}
where $k_b$ is the Boltzmann constant. As can be seen, low frequencies (small $\omega_0$) dramatically increase the Kerr-like response of the superconductor. So, the red shift in the driving pulse, in turn, more effectively melts the superconductor and dynamically reinforces the melting process in a positive feedback loop, as alluded to above.
 This inverse plasma redshift intuitively explains the giant redshift~\cite{robson_giant_2018} and why the redshifted pulse is linked to more substantial melting. 

Next, we consider the role of the chirp explicitly. That the sign of chirp dramatically affects the degree of broadening induced by nonlinear processes has long been known for conventional self-phase modulation~\cite{10.1063/1.109820}, with the exact behavior depending on whether the newly generated frequencies are additive or out-of-phase with existing spectral components of the pulse. However, in the case of a melting superconductor, there is an additional history effect, as the magnitude of $\chi^{(3)}$ is directly related to the number of cooper pairs that remain in the sample, as represented by the $(T-T_c)^2$ factor in Eq.~\eqref{eq nonlinear electric susceptibility}. A positively chirped pulse thus red-shifts the low-frequency components effectively, leading to very strong superconducting suppression in the leading edge of the pulse but comparatively little red-shifting in the trailing edge. As such, only the leading edge effectively melts the superconductor, as can be seen examining Fig.~\ref{fig:c}, leading to the most substantial absorption at these frequencies. In contrast, the trailing high-frequency edge undergoes only a minor redshift. 

In contrast, for a negative chirped pulse, the redshift is strongest for the high-frequency components, and the overall melting process is more distributed over the entire pulse duration, as seen in Fig.~\ref{fig:b}. This leads to a more even shift to low frequencies across the whole bandwidth and allows the total spectral bandwidth to contribute to the suppression of the superconductivity. Thus, this history effect explains why both the red-shift of negatively chirped pulses and the narrowing of positively chirped pulses saturate so effectively in Fig.~\ref{Fig_spectras}, and why the effects depend intimately on the particular nature of the melting of the superconducting condensate.

\section{Conclusion}
With increasingly high-field terahertz pulses being used to manipulate superconductors, the back-reaction of the superconductor upon the driving pulse will become more and more significant. Here, we have presented a study of the entire field and material evolution of a BCS-type superconductor driven by high-field strength THz pulses through a clear and transparent approach based on the TDGL theory. We find the dynamics of the superconductor to be strongly dependent on the initial chirp of the driving pulse, leading to variations up to one order of magnitude of the melting of the superconductor and large-scale spectral shifting of the driving pulse within a few hundred nanometers of propagation through a superconductor. We attribute these dynamics to an inverse plasma redshift response of the condensate. Similarly to the plasma-blueshift phenomena known in plasma physics, here the destruction of Cooper pairs caused by the driving field effectively decreases the free-carrier density/mobility, causing a time-dependent change of the refractive index, which red-shifts the driving field, leading to stronger non-thermal suppression of the condensate. In the limit of strong melting, this shift depends upon the history of the pulse, so the input chirp can modulate the overall degree of melting significantly. 

This history effect and giant red-shift are rather unique features of superconductors. Thus, the uncovered high dependence of the state of the superconductor on purely dynamical variables, such as the chirp, indicates that pulse shaping could be an interesting tool for the control and diagnosis of superconductivity. It also suggests that superconductors could form an interesting class of non-linear optic materials in the THz regime. Our simulations show that macroscopic models of the superconductor-light interaction can reveal unexpected effects in the dynamics of such systems and may need to be explicitly treated to understand experiments performed at high field strengths. \\

\section{Acknowledgements.}

This work was funded by the Spanish AIE (projects PID2022-137817NA-I00 and EUR2022-134052). 

ICFO group acknowledges support from:
ERC AdG NOQIA; MCIN/AEI (PGC2018-0910.13039/501100011033,  CEX2019-000910-S/10.13039/501100011033, Plan National FIDEUA PID2019-106901GB-I00, Plan National STAMEENA PID2022-139099NB-I00 project funded by MCIN/AEI/10.13039/501100011033 and by the “European Union NextGenerationEU/PRTR" (PRTR-C17.I1), FPI); QUANTERA MAQS PCI2019-111828-2);  QUANTERA DYNAMITE PCI2022-132919 (QuantERA II Programme co-funded by European Union’s Horizon 2020 program under Grant Agreement No 101017733),  Ministry for Digital Transformation and of Civil Service of the Spanish Government through the QUANTUM ENIA project call - Quantum Spain project, and by the European Union through the Recovery, Transformation and Resilience Plan - NextGenerationEU within the framework of the Digital Spain 2026 Agenda; Fundació Cellex; Fundació Mir-Puig; Generalitat de Catalunya (European Social Fund FEDER and CERCA program, AGAUR Grant No. 2021 SGR 01452, QuantumCAT \ U16-011424, co-funded by ERDF Operational Program of Catalonia 2014-2020); Barcelona Supercomputing Center MareNostrum (FI-2023-1-0013); EU Quantum Flagship (PASQuanS2.1, 101113690, funded by the European Union. Views and opinions expressed are, however, those of the author(s) only and do not necessarily reflect those of the European Union or the European Commission.  Neither the European Union nor the granting authority can be held responsible for them); EU Horizon 2020 FET-OPEN OPTOlogic (Grant No 899794); EU Horizon Europe Program (Grant Agreement 101080086 — NeQST), results incorporated in this standard have received funding from the European Innovation Council and SMEs Executive Agency under the European Union’s Horizon Europe programme)., ICFO Internal “QuantumGaudi” project; European Union’s Horizon 2020 program under the Marie Sklodowska-Curie grant agreement No 847648;  “La Caixa” Junior Leaders fellowships, La Caixa” Foundation (ID 100010434): CF/BQ/PR23/11980043. Views and opinions expressed are, however, those of the author(s) only and do not necessarily reflect those of the European Union, European Commission, European Climate, Infrastructure and Environment Executive Agency (CINEA), or any other granting authority.  Neither the European Union nor any granting authority can be held responsible for them.

ASJ acknowledges the support of the Ramón y Cajal Program (Grant RYC2021-032392-I). IMDEA Nanociencia acknowledges support from the “Severo Ochoa” Programme for Centers of Excellence in R\&D (MICIN, CEX2020-001039-S).

\newpage
\onecolumngrid
\section{Supplemental material \label{sec:SM}}

\begin{figure*} [h]

 \includegraphics[width=0.6\textwidth]{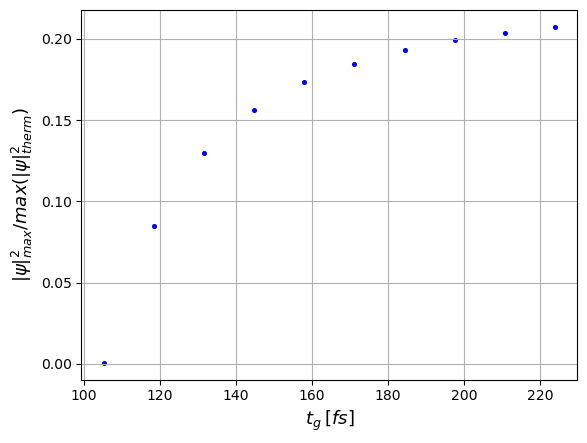}

    \caption{\textbf{Maximum value of the order parameter as a function of the pulse duration}  Maximum value of the order parameter after the pulse has gone through the superconductor, at $t = \SI{12}{\pico\second}$, for incident unchirped pulses with different values of $t_g$. The value is normalized by the maximum value of the initial state, $|\psi|^2_{\text{max}}/\text{max}(|\psi_{\text{therm}}|^2)$, where $\text{max}(|\psi_{\text{therm}}|^2) =  \SI{3.8517}{\per\nano\meter\cubed}$. }
  \label{Fig_suppelmentary}
\end{figure*}

For low-chirp values, the pulse has high-intensity peaks, and our results thus suggest this results in additional non-linear melting processes of the condensate becoming significant. To verify this assertion, we performed additional scans under zero chirp conditions. We keep the total energy in the pulse constant but change $t_g$, decreasing the peak field. As we observe in Fig.~\ref{Fig_suppelmentary}, lower values of $t_g$, with higher peak intensities, significantly suppress the superconducting state, confirming the hypothesis.

\end{document}